\documentclass[aps,pra,twocolumn,superscriptaddress,showpacs,amsmath,amssymb,amsfonts]{revtex4}

\usepackage{graphicx,bm}

\begin{document}

\title{Fidelity approach to Gaussian transitions}

\author{Yu-Chin Tzeng 
}%
\affiliation{Department of Physics, Tunghai University, Taichung,
Taiwan}

\author{Hsiang-Hsuan Hung 
}%
\affiliation{Department of Physics and Astronomy, University of
Pittsburgh, Pittsburgh, Pennsylvania 15260}

\author{Yung-Chung Chen}
\affiliation{Department of Physics, Tunghai University, Taichung,
Taiwan}

\author{Min-Fong Yang}
\email{mfyang@thu.edu.tw} %
\affiliation{Department of Physics, Tunghai University, Taichung,
Taiwan}

\date{\today}

\begin{abstract}
The fidelity approach to the Gaussian transitions in spin-one
$XXZ$ spin chains with three different values of Ising-like
anisotropy $\lambda$ is analyzed by means of the density matrix
renormalization group (DMRG) technique for systems of large sizes.
We find that, despite the success in the cases of $\lambda=2.59$
and $1$, the fidelity susceptibility fails to detect the Gaussian
transition for $\lambda=0.5$. Thus our results demonstrate the
limitation of the fidelity susceptibility in characterizing
quantum phase transitions, which was proposed recently in general
frameworks.
\end{abstract}


\maketitle

\section{Introduction}
\label{sect_intro}

Due to latest progress in quantum information
science~\cite{Nielsen:book}, people attempt to investigate the
quantum phase transitions (QPTs)~\cite{Sachdev:book} in many-body
systems from the perspective of quantum
information~\cite{Amico0703044}. Recently, it was suggested that
the ground-state fidelity, a basic notion that emerged from
quantum information science, should be able to signal the
QPTs~\cite{HTQuan2006,Zanardi06}. Because the fidelity is a
measure of distinguishability between two states, one anticipates
that the ground-state fidelity typically drops abruptly at the
quantum critical point, as a consequence of the dramatic changes
in the structure of the ground states when systems undergo a QPT.
Thus the fidelity should be able to furnish a signature of the
quantum critical point. Another related transition indicator is
provided by the singularity in the second derivative of the
fidelity~\cite{YLG07,ZGC07,ZZWL0711.4651} (or the so-called
``fidelity susceptibility"~\cite{YLG07}). The
fidelity approach has been tested for various many-body systems%
~\cite{HTQuan2006,Zanardi06,YLG07,ZCG07,Cozzini07,CIZ07,Buonsante07,Oelkers07,%
ZGC07,CVZ07,CWGW07,GKNL0706.2495,zhou,Yang07,Fjaerestad,ZOV08,TB08,ZZWL0711.4651,%
ZZ0803.0814,YGSL0803.1292,YGSL0803.2243,HZHL0705.0026,PV08,PSNVD0708.3494,Tzeng08,%
Ning0708.3178,CWHW08,CamposVenuti0801.2473,finiteT}. Since the
fidelity is a purely Hilbert-space geometrical quantity, no
\textit{a priori} knowledge of the structure (order parameter,
correlations driving the QPTs, topology, etc.) of the considered
system is required for its use. Therefore, the fidelity analysis
might be a potential universal tool for characterizing the QPTs.
That is the reason why the investigation of the role of fidelity
in detecting the topological phase transition, where no local
order parameter can distinguish the ground states around the
critical points, has started quite
recently~\cite{HZHL0705.0026,ZZ0803.0814,YGSL0803.1292,YGSL0803.2243}.

Besides the examination of its validity in restricted situations,
general understanding of the fidelity approach has come out, and
the limitation of the fidelity susceptibility in detecting QPTs
has been established~\cite{CVZ07,Yang07}. Conventionally, QPTs are
characterized by the nonanalyticities of the ground-state energy:
first-order QPTs (1QPTs) are characterized by discontinuity in the
first derivative of the ground-state energy, second-order QPTs
(2QPTs) are characterized by discontinuity/singularity in the
second derivative, and so on. It is pointed out that, while the
fidelity susceptibility is indeed an effective tool in detecting
the critical points of 1QPTs and 2QPTs, it may fail to detect
higher-order QPTs~\cite{Yang07}. Moreover, the authors in
Ref.~\cite{CVZ07} show that, for critical (gapless) systems of
finite size $L$, the fidelity susceptibility ${\cal S}$ fulfills
scaling relations
\begin{equation}
{\cal S}\sim L^{-\Delta_{Q}} \; , \qquad
\Delta_{Q}=2\Delta_{V}-2z-d \; ,  \label{scaling_S}
\end{equation}
where $d$ is the spatial dimension, $z$ is the dynamic exponent,
and $\Delta_{V}$ is the scaling dimension of the
transition-driving term in the Hamiltonian. The scaling relation
in Eq.~(\ref{scaling_S}) implies that, for critical points with
$\Delta_{Q}>0$, ${\cal S}$ is non-singular even in the
thermodynamic limit. Therefore, the fidelity susceptibility is not
suitable to single out these QPTs.

Due to the lack of knowledge of exact ground state wave functions,
the ground-state fidelity is usually difficult to be calculated.
For models that are not exactly solvable, most of researchers
resort to numerical exact diagonalization on small systems.
However, because of finite-size effects, the relevance of their
results for the thermodynamic limit may not be guaranteed.
Therefore, further investigations for systems of large sizes are
clearly called for. By using the density matrix renormalization
group (DMRG) technique~\cite{DMRG}, the fidelity susceptibility of
the spin-one $XXZ$ spin chains with an on-site anisotropic term
for the case of the Ising-like anisotropy parameter $\lambda=1$ is
evaluated for system sizes up to $L=160$~\cite{Tzeng08}. The
scaling relation in Eq.~(\ref{scaling_S}) has been confirmed
numerically for both the Ising and the Gaussian transitions of
this model.

In the present article, we further pursue the fidelity approach by
analyzing the Gaussian transitions of the same model for various
parameters of Ising-like anisotropy up to $L=400$. Because the
critical exponents of the Gaussian transitions change continuously
along the critical lines, the QPTs of different order can be
realized simply by choosing different Ising-like anisotropy
parameter $\lambda$. Here, we consider three cases of
$\lambda=2.59$, 1, and 0.5, which correspond to 2QPT, third-order
QPT (3QPT), and fifth-order QPT (5QPT), respectively (see
Secs.~\ref{sect_II} and~\ref{sect_III}). We find that, while the
fidelity susceptibility can serve as a valid transition indicator
for the Gaussian transitions in the cases of $\lambda=2.59$ and 1,
it does not show any singularity around the critical point for
$\lambda=0.5$ and fails to detect this higher-order QPT.
Therefore, our results provide a concrete illustration for the
failure of the fidelity susceptibility in characterizing QPTs, and
lend further numerical supports of the general proposals in
Refs.~\cite{CVZ07} and~\cite{Yang07}.

This paper is organized as follows. The model Hamiltonian and its
low-energy effective theory, as well as the general scaling
arguments, are described in Sec.~\ref{sect_II}. Our DMRG
calculations for the fidelity susceptibility are presented in
Sec.~\ref{sect_III}. We summarize and conclude our results in
Sec.~\ref{sect_conclusions}. The appendix contains an outline of
the computation of the ground-state fidelity under DMRG algorithm.

\section{spin-one anisotropic model and its
effective theory} \label{sect_II}

The $S=1$ $XXZ$ chains with uniaxial single-ion-type anisotropy is
defined by the Hamiltonian:
\begin{equation}
H=\sum_{j}\left\{ S_{j}^{x}S_{j+1}^{x} + S_{j}^{y}S_{j+1}^{y} +
\lambda S_{j}^{z}S_{j+1}^{z} + D
\left(S_{j}^{z}\right)^{2}\right\} ,\label{hamilt}
\end{equation}
where $S_j^\alpha$ ($\alpha=x,y,z$) stand for the spin-one
operators at the $j$-th lattice site. $\lambda$ and $D$
parametrize the Ising-like and the uniaxial single-ion
anisotropies, respectively. The full phase diagram consists of six
different phases~\cite{denNijs89,Schulz86} (see
Refs.~\cite{Chen03,DegliEspostiBoschi03,CamposVenuti06-1,CamposVenuti06-2}
for recent numerical determinations). Here we focus our attention
on the Gaussian transitions between the Haldane and the large-$D$
phases. It has been found that the low-energy effective continuum
theory for the Gaussian transitions can be described by the
sine-Gordon model~\cite{DegliEspostiBoschi03,CamposVenuti06-1}
(here we follow the notations used in
Ref.~\cite{CamposVenuti06-1})
\begin{equation}
H_{SG}=
\frac{1}{2}\left[\Pi^{2}+\left(\partial_{x}\Phi\right)^{2}\right]
-\frac{\mu}{a^{2}}\cos\left(\sqrt{4\pi K}\Phi\right) \; ,
\label{eq:SG}
\end{equation}
where $\Pi$ and $\Phi$ are the conjugate bosonic phase fields, and
$a$ is a short-distance cut-off of the order of the lattice
spacing. The coefficient $\mu\propto(D-D_c)$ in the vicinity of
the critical point $D_c$ for a given $\lambda$, and thus becomes
zero at the transition point. The value of the Luttinger liquid
parameter $K$ varies continuously between 1/2 and 2 along the
critical line. Because all the scaling dimensions and the critical
exponents are determined by a single parameter $K$, they also
change continuously along the critical line. From the sine-Gordon
theory~\cite{lukyanov97}, it is found that the critical exponent
of the correlation length
\begin{equation}
\nu=\frac{1}{2-K} \label{eq:nu}
\end{equation}
and the scaling dimension $\Delta_{V}=K$ for the
transition-driving term $\cos(\sqrt{4\pi K}\Phi)$. Thus the
exponent in the scaling relation of the fidelity susceptibility in
Eq.~(\ref{scaling_S}) becomes ($d=1$ and $z=1$ are assumed here)
\begin{equation}
\Delta_{Q}=2K-3 \; . \label{eq:Delta_Q}
\end{equation}
From this expression, one realizes immediately that, for critical
points with $K>3/2$, $\Delta_{Q}$ is positive and then ${\cal S}$
becomes non-singular even in the thermodynamic limit. That is, the
fidelity susceptibility fails to single out the Gaussian
transitions with $K>3/2$.

According to the conventional classification, a 1QPT is
characterized by a finite discontinuity in the first derivative of
the ground state energy. Similarly, a 2QPT is characterized by a
finite discontinuity, or divergence, in the second derivative of
the ground state energy, assuming the first derivative is
continuous. From the effective theory in Eq.~(\ref{eq:SG}) and the
scaling hypothesis, the first derivative of the ground-state
energy density $e(D)$ with respect to $D$ at a fixed $\lambda$
gives~\cite{CamposVenuti06-1}
\begin{eqnarray}
\frac{\partial e(D)}{\partial D} &\sim& \left\langle
\cos\left(\sqrt{4\pi K}\Phi\right)\right\rangle \nonumber \\
&\sim& \textrm{sgn}\left(D-D_c\right)\left|D-D_c\right|^{\rho} ,
\label{eq:1diff}
\end{eqnarray}
where
\begin{equation}
\rho=\nu\Delta_{V}=\frac{K}{2-K} \; . \label{eq:rho}
\end{equation}
From Eq.~(\ref{eq:1diff}), the second derivative of $e(D)$ can be
obtained as
\begin{equation}
\frac{\partial^2 e(D)}{\partial D^2} \sim |D-D_c|^{\rho-1} \; .
\label{eq:2diff_1}
\end{equation}
Therefore, the second derivative of $e(D)$ shows a divergence for
$0<\rho<1$ (or $0<K<1$). In this case, the Gaussian transition
corresponds to a 2QPT. Similarly, for $1<\rho<2$ (or $1<K<4/3$),
the Gaussian transition can be classified as a 3QPT. Finally, as
mentioned above, for the Gaussian transitions with $K>3/2$, the
fidelity susceptibility fails to detect them. In these cases,
their values of $\rho$ will be greater than three, and these
transitions will be identified as the QPTs of higher than forth
order.

\section{DMRG Results}
\label{sect_III}

In the present work, we consider three cases of $\lambda=2.59$, 1,
and 0.5, whose Luttinger liquid parameter has been found to be
$K=0.85$, 1.328, and 1.580,
respectively~\cite{DegliEspostiBoschi03,CamposVenuti06-2}. From
the discussions in the previous section, the QPTs of these three
cases should correspond to 2QPT, 3QPT, and 5QPT, respectively.
Therefore, one expects that the fidelity susceptibility will not
show singularity for the case of $\lambda=0.5$ only.

In the following, our DMRG results are presented in order. The
fidelity susceptibility for system in Eq.~(\ref{hamilt}) of size
$L$ is calculated by~\cite{Cozzini07,Buonsante07}
\begin{equation}
{\cal S}(D)= \lim_{\delta \to 0} \frac{2 [ 1 - {\cal
F}(D,D+\delta)]}{L\;\delta^2} \; , \label{eq:fide_2_deri}
\end{equation}
where the ground-state fidelity is given by~\cite{Zanardi06}
\begin{equation}
{\cal F}(D, D+\delta) = |\langle\Psi_0(D)|\Psi_0(D+\delta)\rangle|
\label{eq:fide_def}
\end{equation}
with $|\Psi_0 (D)\rangle$ and $|\Psi_0 (D+\delta)\rangle$ being
two normalized ground states corresponding to neighboring
Hamiltonian parameters. Details on the computation of the
ground-state fidelity under DMRG algorithm is explained in the
appendix. In our calculations, $\delta=10^{-3}$ is used. Our
results are evaluated by means of the finite-system DMRG technique
under open boundary conditions for system sizes up to $L=400$,
where 300 states per block are kept and five DMRG sweeps are
performed for the truncation error being $10^{-9}$ at most.
Finally, in order to select a specific ground state and stabilize
our calculations of the fidelity around the N\'{e}el phase, a
Zeeman term $h_1 S^z_1$ acting only upon the spin at site 1 is
added in the Hamiltonian of Eq.~(\ref{hamilt}). Here we take the
boundary magnetic field $h_1=-1$.

\subsection{case of $\lambda=2.59$}

\begin{figure}
\includegraphics[width=3.7in]{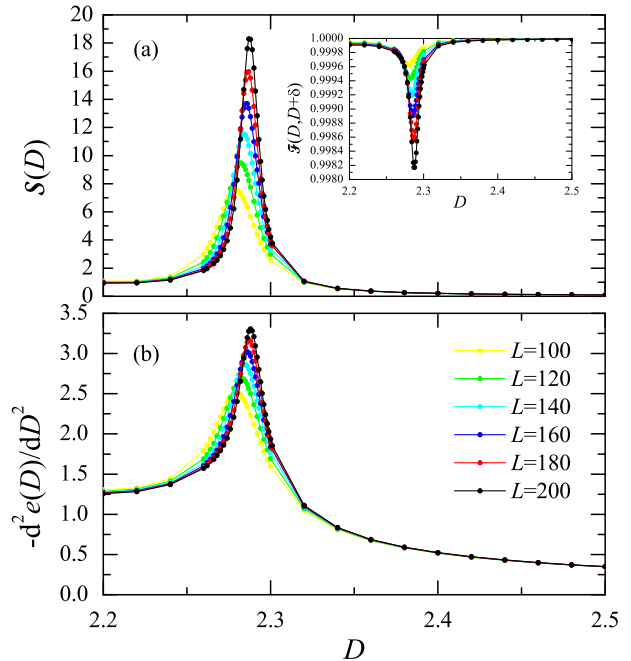}
\vskip -0.5cm%
\caption{(Color online) (a) Fidelity susceptibility ${\cal S}$,
(b) the second derivative of the ground-state energy density
$-\partial^2 e/\partial D^2$ as functions of $D$ for various sizes
$L$ with $\lambda=2.59$. Inset in the top panel shows the fidelity
${\cal F}(D, D+\delta)$ as functions of $D$ for the corresponding
sizes. Here we take $\delta=10^{-3}$.} \label{fig:jz2.59}
\end{figure}

\begin{figure}
\includegraphics[width=3.7in]{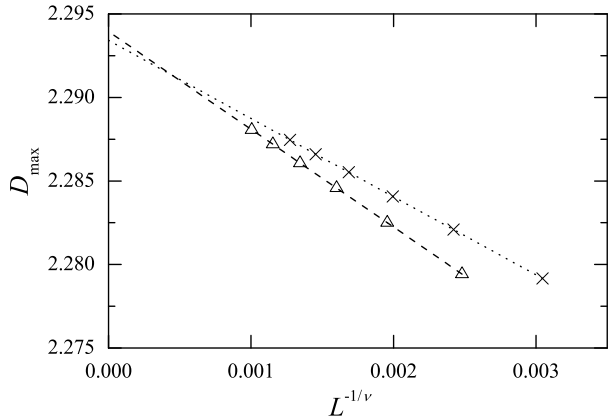}
\vskip -0.5cm%
\caption{Finite-size scaling of $D_{\rm max}$ of ${\cal S}$
($\times$) and $-\partial^2 e/\partial D^2$ ($\triangle$) versus
$L^{-1/\nu}$ for $\lambda=2.59$. The lines are least square fits,
where $\nu\simeq 0.79$ ($\nu\simeq 0.77$) for those $D_{\rm
max}$'s corresponding to the local maxima in the curves of ${\cal
S}$ ($-\partial^2 e/\partial D^2$). } \label{fig:Dc_jz2.59}
\end{figure}

The findings of the fidelity susceptibility ${\cal S}(D)$ and the
ground-state fidelity ${\cal F}(D, D+\delta)$ for $\lambda=2.59$
are shown in Fig.~\ref{fig:jz2.59}(a). As shown in the inset,
drops in the ground-state fidelity are observed, which signal
precursors of the Gaussian transition 
in the model under consideration. Further evidences for indicating
QPTs are provided by the results of the fidelity susceptibility
${\cal S}$ and the second derivative of the ground-state energy
density $\partial^2 e/\partial D^2$. As seen from
Fig.~\ref{fig:jz2.59}, the maximum values of both ${\cal S}$ and
$-\partial^2 e/\partial D^2$ grow with increasing size, and thus
indicate divergence in the $L\to\infty$ limit (see also
Fig.~\ref{fig:Smax_jz2.59} below). As discussed in
Sec.~\ref{sect_II}, the divergent behaviors both in the fidelity
susceptibility and the second derivative of the ground-state
energy density indicate that the transition we found should be a
2QPT.

According to the finite-size scaling theory~\cite{fss}, one has
\begin{equation}
|D_{\rm max}(L) - D_c|  \propto L^{-1/\nu} \; , \label{eq:fss}
\end{equation}
where $D_c$ is the critical point in the thermodynamic limit and
$\nu$ is the critical exponent of the correlation length. Thus
$D_c$ can be determined by an extrapolation procedure from the
locations $D_{\rm max}(L)$ of the local maxima in ${\cal S}$ and
$-\partial^2 e/\partial D^2$ on a size-$L$ system. The results for
$\lambda=2.59$ is shown in Fig.~\ref{fig:Dc_jz2.59}. We find that
$D_c \simeq 2.293$ and $\nu \simeq 0.79$ for the findings of
${\cal S}$, while $D_c \simeq 2.294$ and $\nu \simeq 0.77$ for the
data of $-\partial^2 e/\partial D^2$. Because of the relation in
Eq.~(\ref{eq:nu}), the Luttinger liquid parameter $K\simeq 0.74$
($K\simeq 0.70$) for the data related to ${\cal S}$ ($-\partial^2
e/\partial D^2$). We find that our results are consistent with the
previous findings~\cite{DegliEspostiBoschi03}, where $D_c\simeq
2.30$ and $K\simeq 0.85$. We note that the DMRG calculations in
Refs.~\cite{DegliEspostiBoschi03} and \cite{CamposVenuti06-2} are
under the periodic boundary conditions, rather than the open
boundary conditions used in the present work. The small
discrepancy of our results from theirs may be due to the different
boundary conditions employed.

Moreover, at the critical point, where the only length scale is
provided by the system size itself, the scaling formula in
Eq.~(\ref{eq:2diff_1}) implies
\begin{equation}
\frac{\partial^2 e(D)}{\partial D^2} \sim L^{-(\rho-1)/\nu} \; .
\label{eq:2diff_2}
\end{equation}
From Eqs.~(\ref{eq:nu}) and (\ref{eq:rho}), one has
\begin{equation}
-(\rho-1)/\nu=2(1-K) \; . \label{eq:exponent}
\end{equation}
On the other hand, the fidelity susceptibility ${\cal S}$ is
proposed to fulfill the scaling relation in Eq.~(\ref{scaling_S}).
To verify the predicted critical scaling behaviors in
Eqs.~(\ref{scaling_S}) and (\ref{eq:2diff_2}), the values ${\cal
S}_{\rm max}$ and $[-\partial^2 e/\partial D^2]_{\rm max}$ of the
local maxima in ${\cal S}$ and $-\partial^2 e/\partial D^2$ for
various sizes $L$ are plotted in Figs.~\ref{fig:Smax_jz2.59}(a)
and (b), respectively. It is found that our data do fulfill the
predicted critical scaling behaviors with $\Delta_{Q} \simeq
-1.28$ and $-(\rho-1)/\nu\simeq 0.40$. From
Eqs.~(\ref{eq:Delta_Q}) and (\ref{eq:exponent}), our values of
$\Delta_{Q}$ and $-(\rho-1)/\nu$ gives $K \simeq 0.86$ and $0.80$,
respectively. The Luttinger liquid parameter $K$ determined by the
present finite-size scaling agrees with the previous
findings~\cite{DegliEspostiBoschi03} and those determined by the
critical exponent $\nu$ coming from the scaling in
Fig.~\ref{fig:Dc_jz2.59}. Thus the fact that a single parameter
$K$ controls all the critical exponents for the Gaussian
transition is confirmed by our numerical results.

\begin{figure}
\includegraphics[width=3.7in]{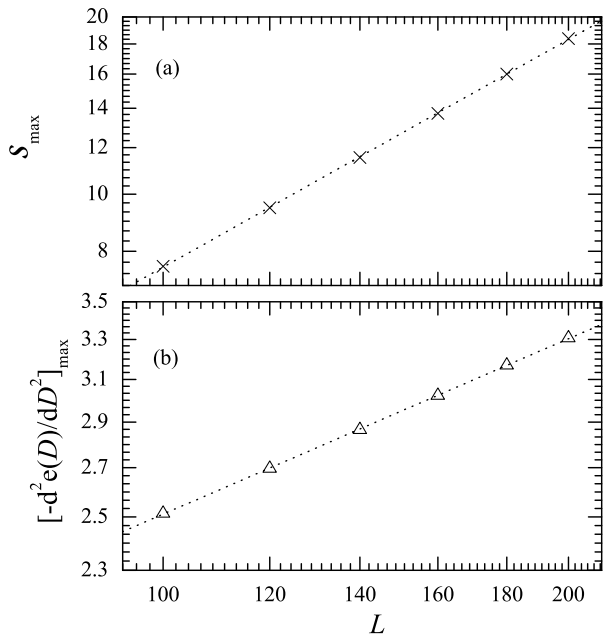}
\vskip -0.5cm%
\caption{The log-log plot of (a) ${\cal S}_{\rm max}$ and (b)
$[-\partial^2 e/\partial D^2]_{\rm max}$ for various sizes $L$
with $\lambda=2.59$. The lines are least square straight line
fits.}%
\label{fig:Smax_jz2.59}
\end{figure}

\subsection{case of $\lambda=1$}

\begin{figure}
\includegraphics[width=3.7in]{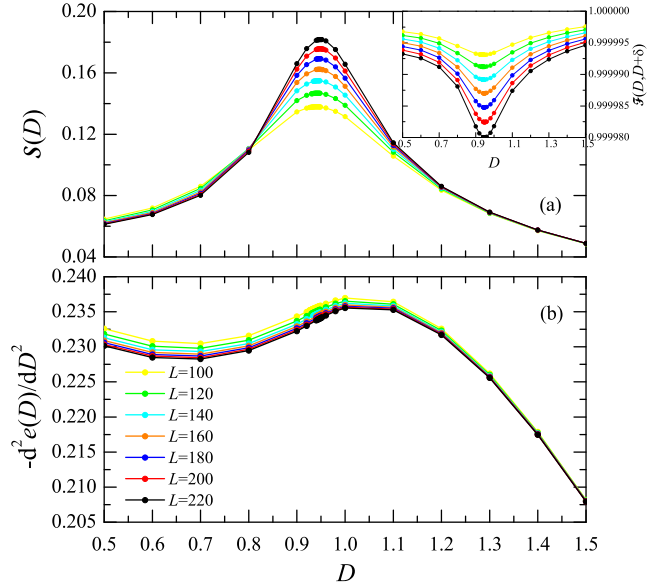}
\vskip -0.5cm%
\caption{(Color online) (a) Fidelity susceptibility ${\cal S}$,
(b) the second derivative of the ground-state energy density
$-\partial^2 e/\partial D^2$ as functions of $D$ for various sizes
$L$ with $\lambda=1$. Inset in the top panel shows the fidelity
${\cal F}(D, D+\delta)$ as functions of $D$ for the corresponding
sizes. Here we take $\delta=10^{-3}$.} \label{fig:jz1}
\end{figure}

For the case of $\lambda=1$, a numerical support of the use of the
fidelity in detecting QPTs has been provided in
Ref.~\cite{Tzeng08}, and the scaling relation of the fidelity
susceptibility in Eq.~(\ref{scaling_S}) has also been verified
there. Here, we focus our attention on the comparison between the
scaling behaviors of the fidelity susceptibility and the second
derivative of the ground-state energy density, which is shown in
Fig.~\ref{fig:jz1}. As observed in Ref.~\cite{Tzeng08}, our
present data in Fig.~\ref{fig:jz1}(a) do show the developing drops
in the ground-state fidelity and the divergent behaviors in ${\cal
S}$. Thus the fidelity susceptibility does signal precursors of
the Gaussian transition in the model under consideration. However,
as seen from Fig.~\ref{fig:jz1}(b), the second derivative of the
ground-state energy density does not grow with increasing size in
the same parameter regime. It implies that this QPT should be the
one of higher than second order. To the author's knowledge, the
present model at $\lambda=1$ may be the first concrete example in
the literature with a QPT of higher than second order, which can
be singled out unambiguously by using the fidelity susceptibility.

\subsection{case of $\lambda=0.5$}

\begin{figure}
\includegraphics[width=3.7in]{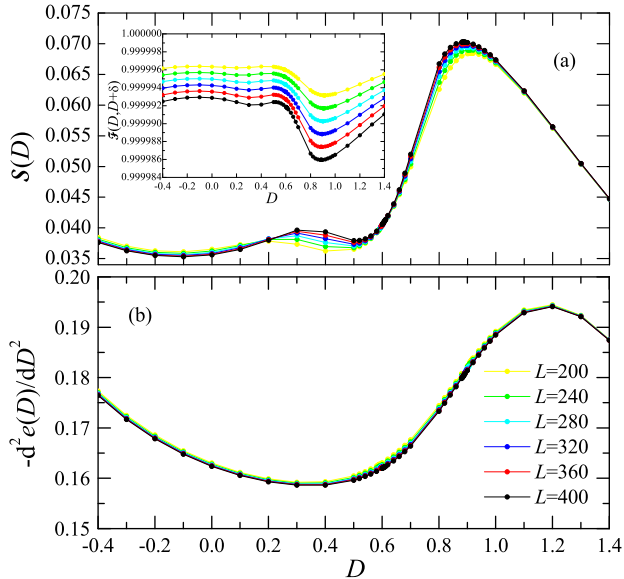}
\vskip -0.5cm%
\caption{(Color online) (a) Fidelity susceptibility ${\cal S}$,
(b) the second derivative of the ground-state energy density
$-\partial^2 e/\partial D^2$ as functions of $D$ for various sizes
$L$ with $\lambda=0.5$. Inset in the top panel shows the fidelity
${\cal F}(D, D+\delta)$ as functions of $D$ for the corresponding
sizes. Here we take $\delta=10^{-3}$.} \label{fig:jz0.5}
\end{figure}

Although the fidelity susceptibility is possible to detect the
QPTs of higher than second order, as explained in
Ref.~\cite{Yang07}, it may not always work. For example, as
discussed at the end of Sec.~\ref{sect_II}, the fidelity
susceptibility will fail to detect those Gaussian transitions with
the Luttinger liquid parameter $K>3/2$, which give $\Delta_{Q}>0$
and correspond to the QPTs of higher than forth order. In the
following, we show that the present model at $\lambda=0.5$
provides an example for this case.

\begin{figure}
\includegraphics[width=3.7in]{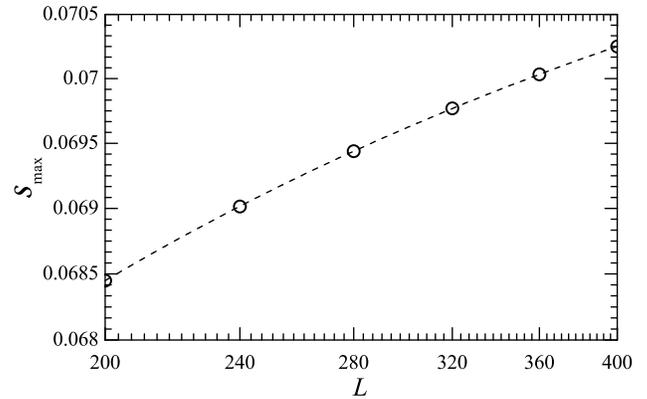}
\vskip -0.5cm%
\caption{The log-log plot of ${\cal S}_{\rm max}$ for various
sizes $L$ with $\lambda=0.5$. The dashed curve is the least square
fit by using the formula ${\cal S}_{\rm max}={\cal
S}_{\infty}-aL^{-b}$ with ${\cal S}_{\infty}=0.073$, $a=0.23$, and
$b=0.75$.}%
\label{fig:Smax_jz0.5}
\end{figure}

Our DMRG results of the fidelity susceptibility ${\cal S}(D)$, the
ground-state fidelity ${\cal F}(D, D+\delta)$, and  the second
derivative of the ground-state energy density $\partial^2
e/\partial D^2$ for $\lambda=0.5$ are shown in
Fig.~\ref{fig:jz0.5}. While developing drops in the ground-state
fidelity are indeed observed in the inset of
Fig.~\ref{fig:jz0.5}(a), the fidelity susceptibility and the
second derivative of the ground-state energy density show no
divergent behavior up to $L=400$. That is, this Gaussian
transition must be a QPT of higher than second order, which the
fidelity susceptibility fails to detect. Further support for such
a conclusion is provided by the findings of ${\cal S}_{\rm max}$
of the local maxima in ${\cal S}$ for various sizes $L$, which is
plotted in Fig.~\ref{fig:Smax_jz0.5}. It is found that maximal
values of ${\cal S}$ do tend to saturate, rather than diverge, in
the $L\to\infty$ limit. Thus this Gaussian transition does provide
a counterexample of the usefulness of the fidelity susceptibility
as an effective tool in detecting QPTs.

\begin{figure}
\includegraphics[width=3.7in]{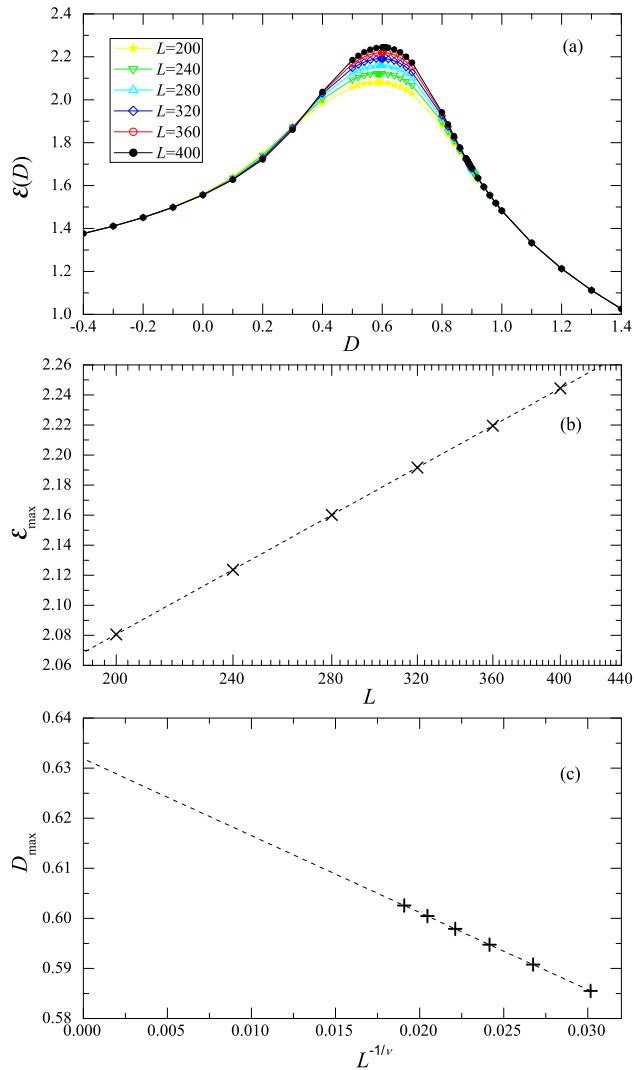}
\vskip -0.5cm%
\caption{(Color online) (a) Entanglement entropy ${\cal E}$ as
functions of $D$ for various sizes $L$. (b) The semi-log plot of
${\cal E}_{\rm max}$ for various sizes $L$. The line is least
square straight line fit. (c) Finite-size scaling of $D_{\rm max}$
of ${\cal E}$ versus $L^{-1/\nu}$. The line is least square
fit with $\nu\simeq 1.51$. Here $\lambda=0.5$.}%
\label{fig:EE_jz0.5}
\end{figure}

One may wonder if the non-singular behavior in the fidelity
susceptibility shown in Figs.~\ref{fig:jz0.5} and
\ref{fig:Smax_jz0.5} is due to the possible finite-size effects,
even though sizes up to $L=400$ have been considered. By contrast,
the entanglement entropy is evaluated below. It has been found
that the divergent character of the entanglement entropy can also
show the existence of the
QPTs~\cite{Vidal03,Holzhey,Korepin04,CC04,LSCA,ZBFS,Cardy0708.2978}.
Within the present DMRG calculations, the singular behavior of the
entanglement entropy is indeed observed in the given parameter
regime (see Fig.~\ref{fig:EE_jz0.5} below). Thus the desired
Gaussian transition does appear there. This indicates that the
system sizes we employed should be large enough and finite-size
effects should play no important role in our calculations.

Here we consider the entanglement entropy, or the von Neumann
entropy of the reduced density matrix $\rho_R(D)$ of the
right-hand block of $L/2$ contiguous spins
\begin{equation}
{\cal E}(D) = -{\rm Tr} \left[ \rho_R(D) {\rm log}_2 \rho_R(D)
\right] \; . \label{eq:entropy}
\end{equation}
Our DMRG results are presented in Fig.~\ref{fig:EE_jz0.5}. It is
found from Fig.~\ref{fig:EE_jz0.5}(a) that peaks do grow as size
$L$ increases, which show the existence of the expected Gaussian
transition. Further evidence is provided by the findings of ${\cal
E}_{\rm max}$ of the local maxima in ${\cal E}$ for various sizes
$L$, which is plotted in Fig.~\ref{fig:EE_jz0.5}(b). It indicates
that ${\cal E}_{\rm max}\propto \log_2(L)$ and thus ${\cal E}_{\rm
max}$ will diverge in the $L\to\infty$ limit. It is known that the
slope in Fig.~\ref{fig:EE_jz0.5}(b) gives the value of $c/6$ under
open boundary
conditions~\cite{Vidal03,Holzhey,Korepin04,CC04,LSCA,ZBFS,Cardy0708.2978},
where $c$ is the central charge of the conformal field theory. Our
data give $c\simeq 0.98$, which agrees with the theoretical value
$c=1$ for the Gaussian transition~\cite{DegliEspostiBoschi03}.
Furthermore, according to Eq.~(\ref{eq:fss}), the corresponding
critical points $D_c$ can be deduced from the locations $D_{\rm
max}(L)$ of the local maxima in ${\cal E}(D)$ on a size-$L$
system, as shown in Fig.~\ref{fig:EE_jz0.5}(c). It is found that
$D_c\simeq 0.63$ and $\nu \simeq 1.51$. The value of $\nu$ implies
the Luttinger liquid parameter $K\simeq 1.34$ due to the relation
in Eq.~(\ref{eq:nu}). In Ref.~\cite{DegliEspostiBoschi03},
$D_c\simeq 0.65$ and $K\simeq 1.580$ are reported. Thus our
results are again consistent with their findings. As mentioned
before, the discrepancy of our results from theirs may be due to
the different boundary conditions employed. In short, the missing
QPT in the measurements of the fidelity susceptibility is able to
be captured by using the tool of the entanglement entropy.

\section{Conclusions}
\label{sect_conclusions}

In summary, we present numerical analysis of the fidelity approach
to the Gaussian transitions in the spin-one $XXZ$ spin chains in
Eq.~(\ref{hamilt}) with three different values of Ising-like
anisotropy $\lambda$. It is found that the fidelity susceptibility
can detect the Gaussian transitions for the cases of
$\lambda=2.59$ and $1$. As in the case of $\lambda=1$ investigated
in Ref.~\cite{Tzeng08}, the scaling relation of the fidelity
susceptibility proposed in Ref.~\cite{CVZ07} is again verified by
the present DMRG calculations for the case of $\lambda=2.59$.
Moreover, the critical point and some of the corresponding
critical exponents are determined through a proper finite-size
scaling analysis, and these values agree with the findings
reported in the literature. Although it is successful in
characterizing the QPTs for the cases of $\lambda=2.59$ and $1$,
the fidelity susceptibility is not able to detect the Gaussian
transitions for $\lambda=0.5$. This observation can be explained
by the general theories in Refs.~\cite{CVZ07} and \cite{Yang07}.
Thus our work provides an instance for the limitation of the use
of the fidelity susceptibility for revealing QPTs.

Finally, we remind that developing drops in the ground-state
fidelity are observed around the critical point even for the case
of $\lambda=0.5$ [see the inset of Fig.~\ref{fig:jz0.5}(a)]. This
seems to indicate that the singularity of the fidelity may still
be a valid transition indicator even for higher-order QPTs. The
failure of its second derivatives (say, the fidelity
susceptibility) may just imply that
higher derivatives of the fidelity are necessary to signal some
higher-order QPTs.
Indeed, such a conclusion can be deduced from the general
arguments in Refs.~\cite{CVZ07} and \cite{Yang07}. It was shown in
Ref.~\cite{Yang07} that the fidelity susceptibility ${\cal S}$ (or
the second derivative of the fidelity per site) can be associated
with the first derivative of the density matrix $\rho_0$ of the
ground state [see Eq.~(6) therein]. Thus ${\cal S}$ can not detect
the higher-order QPTs resulting from the nonanalyticities in
second (and even higher-order) derivatives of $\rho_0$. That is,
these higher-order QPTs can be recognized only by the singular
behavior in higher-order derivatives of the fidelity. Besides, the
singular part of ${\cal S}$ has been shown in Ref.~\cite{CVZ07} to
behave as ${\cal S}\sim |D-D_c|^{\nu\Delta_{Q}}$. Therefore, the
third derivative of the fidelity should behave as
$|D-D_c|^{\nu\Delta_{Q}-1}$. This implies that, although the
second derivative of the fidelity (or the fidelity susceptibility
${\cal S}$) does not diverge for the Gaussian transitions with
$3/2<K<5/3$ [thus $0<\nu\Delta_{Q}<1$, see Eqs.~(\ref{eq:nu}) and
(\ref{eq:Delta_Q})], the third derivative does become singular at
these critical points. That is, one does need the third derivative
of the fidelity to single out these higher-order QPTs.
The more complete and deeper understanding of the role played by
the fidelity and its derivatives in the studying of the
higher-order QPTs is an interesting issue which calls for future
investigations.

\begin{acknowledgments}
YCT and YCC are grateful for the support from the National Science
Council of Taiwan under NSC 96-2112-M-029-003-MY3. MFY acknowledge
the support by the National Science Council of Taiwan under NSC
96-2112-M-029-004-MY3.
\end{acknowledgments}

\appendix*
\section{computation of fidelity under DMRG algorithm}

\begin{figure}
\includegraphics[width=3in]{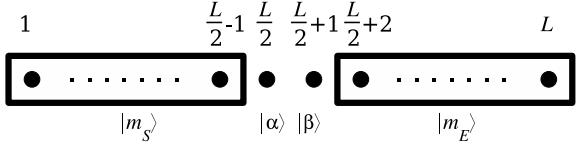}
\caption{The configuration of wave functions in the superblock.}
\label{superblock}
\end{figure}

\begin{figure}
\includegraphics[width=3in]{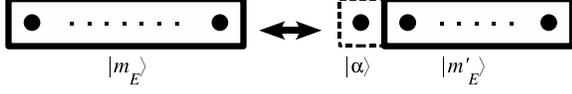}
\caption{The basis transformation between a block and an enlarged
block.} \label{transorm}
\end{figure}

The DMRG method had been used to calculate the overlap of two
different ground states for a decade. For example, people had
applied this technique to evaluate the exponent of the
orthogonality catastrophe for the problem of a single impurity in
a one-dimensional Luttinger liquid~\cite{SJQin}. Recently, this
method is used to investigate the scaling properties of fidelity
susceptibility for the model in Eq.~(\ref{hamilt}) with
$\lambda=1$~\cite{Tzeng08}. For completeness, the algorithm within
the DMRG framework of computation of the ground-state fidelity
(i.e., the overlap between two ground states corresponding to two
different system parameters) is presented in this appendix.
Similar discussions in terms of the formalism of matrix product
states can be found at the end of Sec.~2.1 in
Ref.~\cite{IPMcCulloch}.

Let $|\Phi\rangle$ and $|\Psi\rangle$ be the normalized wave
functions obtained from two different DMRG calculations with
distinct Hamiltonian parameters for systems of the same size.
Assume that the same number of states $m$ is kept in both DMRG
calculations. Moreover, both wave functions are expressed in the
same scheme, where the corresponding system (and thus the
corresponding environment) blocks are of the same length.
Therefore, these two wave functions can be written as (see
Fig.~\ref{superblock})
\begin{equation}
|\Psi\rangle = \sum_{m_S,\alpha,\beta,m_E}
\Psi_{m_S,\alpha,\beta,m_E} |m_S\rangle
|\alpha\rangle |\beta\rangle |m_E\rangle \; ,
\end{equation}
\begin{equation}
|\Phi\rangle =
\sum_{\bar{m}_S,\alpha^\prime,\beta^\prime,\bar{m}_E}
\Phi_{\bar{m}_S,\alpha^\prime,\beta^\prime,\bar{m}_E}
|\bar{m}_S\rangle |\alpha^\prime\rangle |\beta^\prime\rangle
|\bar{m}_E\rangle \; .
\end{equation}
Here $|\alpha\rangle$ ($|\alpha^\prime\rangle$) and
$|\beta\rangle$ ($|\beta^\prime\rangle$) denote the states in the
local Hilbert space of dimension $s$ at sites $L/2$ and $L/2+1$,
respectively. $|m_S\rangle$ and $|\bar{m}_S\rangle$ represent the
$m$ renormalized states of the system blocks, while $|m_E\rangle$
and $|\bar{m}_E\rangle$ are those $m$ renormalized states of the
environment blocks. Thus the overlap between these two ground
states becomes
\begin{widetext}
\begin{equation}\label{overlap-1}
\langle\Phi|\Psi\rangle =
\sum_{\bar{m}_S,\alpha^\prime,\beta^\prime,\bar{m}_E;m_S,\alpha,\beta,m_E}
\Phi^*_{\bar{m}_S,\alpha^\prime,\beta^\prime,\bar{m}_E}
\langle\bar{m}_S|m_S\rangle \langle\alpha^\prime|\alpha\rangle
\langle\beta^\prime|\beta\rangle \langle\bar{m}_E|m_E\rangle
\Psi_{m_S,\alpha,\beta,m_E}  \; .
\end{equation}
\end{widetext}
Remind that $\langle\alpha^\prime|\alpha\rangle$ and
$\langle\beta^\prime|\beta\rangle$ are nothing but the matrix
elements of the $s\times s$ identity matrix $\hat{1}$ in the
subspaces of the local Hilbert spaces at sites $L/2$ and $L/2+1$,
respectively. Besides, let's define
\begin{eqnarray}
\protect[\hat{1}_{\frac{L}{2}-1}^S\protect]_{\bar{m}_S,m_S} &=&
\langle\bar{m}_S|m_S\rangle \; ,  \\
\protect[\hat{1}_{\frac{L}{2}-1}^E\protect]_{\bar{m}_E,m_E} &=&
\langle\bar{m}_E|m_E\rangle \; ,\label{def_E}
\end{eqnarray}
which serve as the matrix elements of the $m\times m$ ``identity
matrices" $[\hat{1}_{\frac{L}{2}-1}^S]$ and
$[\hat{1}_{\frac{L}{2}-1}^E]$ of the system and the environment
blocks of length $\frac{L}{2}-1$, respectively. Therefore, the
expression of the overlap in Eq.~(\ref{overlap-1}) can be
rewritten as the following compact from
\begin{equation}\label{overlap}
\langle\Phi|\Psi\rangle = \langle\langle\Phi|| \, [\hat{1}] \,
||\Psi\rangle\rangle
\end{equation}
with
\begin{equation}
[\hat{1}] = [\hat{1}_{\frac{L}{2}-1}^S] \otimes \hat{1} \otimes
\hat{1} \otimes [\hat{1}_{\frac{L}{2}-1}^E] \; ,
\end{equation}
where $||\Phi\rangle\rangle$ and $||\Psi\rangle\rangle$ represent
the column vectors of dimension $s^2m^2$ with the components
$\Phi_{\bar{m}_S,\alpha^\prime,\beta^\prime,\bar{m}_E}$ and
$\Psi_{m_S,\alpha,\beta,m_E}$, respectively. Now the remaining
work of evaluation of fidelity reduces to find a general strategy
of calculating the two $m\times m$ matrices
$[\hat{1}_{\frac{L}{2}-1}^S]$ and $[\hat{1}_{\frac{L}{2}-1}^E]$.

Note that the renormalized states of the environment blocks of
length $l$ can be expressed as (see Fig.~\ref{transorm})
\begin{eqnarray}
|m_E\rangle &=& \sum_{\alpha,m^\prime_E}
[O^E_l]_{(\alpha,m^\prime_E),m_E}
|\alpha\rangle |m^\prime_E\rangle  \; , \label{right_transform-1} \\
|\bar{m}_E\rangle &=& \sum_{\beta,\bar{m}^\prime_E}
[\bar{O}^E_l]_{(\beta,\bar{m}^\prime_E),\bar{m}_E} |\beta\rangle
|\bar{m}^\prime_E\rangle \; , \label{right_transform-2}
\end{eqnarray}
where $[O^E_l]_{(\alpha,m^\prime_E),m_E}$ and
$[\bar{O}^E_l]_{(\beta,\bar{m}^\prime_E),\bar{m}_E}$ denote the
matrix elements of an $sm\times m$ transformation matrices
$[O^E_l]$ and $[\bar{O}^E_l]$ {\it for two different DMRG
calculations with distinct system parameters}. They are formed by
$m$ eigenvectors with the largest eigenvalues of the reduced
density matrix of the environment blocks of length $l$. By using
Eqs.~(\ref{right_transform-1}) and (\ref{right_transform-2}), one
obtains
\begin{widetext}
\begin{equation}\label{state_transform-1}
\langle\bar{m}_E|m_E\rangle =
\sum_{(\beta,\bar{m}^\prime_E);(\alpha,m^\prime_E)}
\left([\bar{O}^E_l]_{(\beta,\bar{m}^\prime_E),\bar{m}_E}\right)^*
\langle\beta|\alpha\rangle
\langle\bar{m}^\prime_E|m^\prime_E\rangle
[O^E_l]_{(\alpha,m^\prime_E),m_E} \; .
\end{equation}
\end{widetext}
Following the similar definition in Eq.~(\ref{def_E}), the above
expression can be casted into a compact form:
\begin{equation}\label{El}
[\hat{1}_l^E]=[\bar{O}^E_l]^{\dagger} \left(\hat{1} \otimes
[\hat{1}_{l-1}^E]\right)[O^E_l] \; ,
\end{equation}
which gives the recursive relations of the $m\times m$ ``identity
matrices" $[\hat{1}_l^E]$ of the environment block of length $l$.
For instance, in the cases of $l=2$ and 3, one has
\begin{equation*}
[\hat{1}_2^E]=[\bar{O}^E_2]^{\dagger}\left( \hat{1} \otimes\hat{1}
\right)[O^E_2] \; ,
\end{equation*}
\begin{equation*}
[\hat{1}_3^E]=[\bar{O}^E_3]^{\dagger}\left(\hat{1}\otimes[\hat{1}_2^E]
\right)[O^E_3] \; .
\end{equation*}
Similar recursive relations for the system blocks reads
\begin{equation}\label{Sl}
[\hat{1}_{l}^S]=[\bar{O}^S_{l}]^{\dagger} \left([\hat{1}_{l-1}^S]
\otimes \hat{1} \right)[O^S_{l}] \; .
\end{equation}
If we iterate all the way from both ends of the systems by
employing the recursive relations in Eqs.~(\ref{El}) and
(\ref{Sl}), then the matrices $[\hat{1}_{\frac{L}{2}-1}^S]$ and
$[\hat{1}_{\frac{L}{2}-1}^E]$ can be constructed eventually.
Substituting these results into Eq.~(\ref{overlap}), the value of
the ground-state fidelity will be obtained.

\end{document}